\input lecproc.cmm
\input epsf.sty

\vsize=23true cm
\hsize=15true cm

%---------------------------------------------------------------------------
\contribution{Average Spectral Energy Distributions of Blazars from 
Radio to $\gamma$--rays}
\contributionrunning{SEDs of Blazars}

\author{G. Fossati@1, A. Celotti@1, G. Ghisellini@2 and L. Maraschi@2 } 
\authorrunning{G. Fossati et al.}
\address{@1International School for Advanced Studies, via Beirut 4, Trieste, Italy
@2Osservatorio Astronomico di Brera, via Brera 28, Milano, Italy}

%---------------------------------------------------------------------------

\abstract{
The average Spectral Energy Distributions (SED) of different classes of 
blazars from the radio to the gamma--ray band are studied adopting a 
``complete--sample" approach.
The SED are globally similar, despite systematic differences in specific 
frequency ranges, showing two broad peaks. However, 
i) the first and second peak occur in different frequency ranges for 
different objects, with a tendency for the most luminous sources to peak at 
lower frequencies, 
ii) the ratio between the two peak frequencies {\it seems to be constant}, 
iii) the luminosity ratio between the high and low frequency component 
increases with the bolometric luminosity.
}

\titleb{1}{The Blazar Family}

The blazars family comprises various kind of sources showing a broad
range of properties, thus raising the question as to up to which degree 
and in which form they are physically related. 
For instance:
i) the properties of BL Lac objects selected in the radio/X--ray spectral 
bands are systematically different, the first difference to be recognized 
and perhaps still the most striking being the shape of the SED.
ii) is there any difference between $\gamma$--ray detected
blazars and the rest of the class ? 

We would like to shed light on these kind of issues adopting an objective 
point of view. 
In the absence of ``blazar complete samples", we consider well defined samples
of blazars, covering a large range of properties.  
Indeed one problem to overcome is that until now the observational approach 
has been mainly ``sub--class oriented'', focusing on either 
BL Lacs or FSRQ/HPQ/OVVs separately.

\noindent
We considered three samples of blazars: 

\smallskip
\item{$\bullet$} the ``1 Jy'' sample of BL Lacs, radio--selected, 34 sources
(Stickel et al. 1991);
\item{$\bullet$} the {\sl Einstein} Slew survey (quasi)--complete sample, X--ray selected, 48 sources (Perlman et al. 1996a);
\item{$\bullet$} the FSRQ complete sample drawn by Padovani and Urry (1992)
from the Wall and Peacock (1985) catalogue of radio--sources with 
F$_{\rm 2.7 GHz} > 2$ Jy, 50 sources.
\smallskip

%% \titlec{2.1}{Radio--X--ray SEDs}
\titleb{2}{The SEDs}

\noindent{\bf $\bullet$ Radio to X--rays: }
average spectral energy distributions from the radio to the X--ray band 
have been constructed. 
A large fraction of the 132 sources have been observed with the {\it ROSAT} 
PSPC allowing to derive "uniformly"  X--ray fluxes and in many cases spectral 
shapes in the 0.1 -- 2.4 keV range (Comastri et al. 1997, Lamer et al. 1996, 
Perlman et al. 1996b, Sambruna et al. 1996a, Sambruna 1997).
Fluxes at radio, mm (230 GHz), and optical frequencies were taken from the 
literature.

The average SEDs presented here are obtained dividing sources in radio
(5 GHz) luminosity bins.
The true goal would be to divide sources according to their 
bolometric luminosity 
(in most cases substantially corresponding to the $\gamma$--ray one).
On the basis of the empirical relationship that seems to hold between 
$\gamma$--ray and radio ({\it e.g.} Mattox 1997) luminosity as a first 
and simpler approach we therefore use the radio power as a good indicator 
of the total intrinsic luminosity.

Within each luminosity bin, averages were performed over the logarithms 
of the K--corrected fluxes at each frequency. 
The average fluxes were then transformed to average luminosities using
the average logarithmic distance of the class.  

%% \titlec{2.2}{$\gamma$--ray spectra}

\smallskip
\noindent{\bf $\bullet$ $\gamma$--ray spectra: }
among the three samples only a fraction of blazars were detected 
in $\gamma$--rays, namely 9/34 in ``1 Jy'' sample, 8/48 in Slew sample,
19/50 in FSRQ sample. 
Many other blazars ($\sim~25$) have been detected by EGRET, but do not fall 
in any well defined sample, apart from the $\gamma$--ray one.

To take advantage of the whole body of information
regarding the $\gamma$--ray properties of blazars, we associated 
$\gamma$--ray properties to our average SEDs following an indirect procedure.
We associated to each averaged SED the ($\langle$L$_{\gamma}\rangle$,
$\langle\alpha_\gamma\rangle$) of blazars taken from the whole EGRET sample, 
falling in the corresponding L$_{\rm 5GHz}$ bin, 
disregarding  their belonging to our complete list. 
The basic assumption is the uniformity of the spectral properties (for a
discussion on this see Impey 1996).

\titleb{3}{Results}

As a first check we compared the distributions of various broad band
spectral indices ({\it e.g.} radio--to--X--ray) and luminosities.
It is striking that there is no apparent difference in any of the
considered quantities between the distributions of $\gamma$--ray detected 
sources and the rest of sample, apart maybe for the Slew sample where 
there seems to be a tendency for $\gamma$--ray detected sources to have 
larger $\alpha_{RX}$ and L$_{\rm 5GHz}$.
This fact can be due to two different and plausible reasons, that in 
any case can not be disentangled: 
i) more luminous sources are more likely to be detected, 
ii) sources with smaller $\alpha_{RX}$ would have the peak of their Compton
component beyond the EGRET band and so they are not easily detectable.

In addition to the average SEDs for each of the three samples (for a
thorough discussion see Fossati et al. 1997b), we considered a
{\it total blazars sample}, resulting from their sum. 
We similarly constructed average SEDs from it, in L$_{\rm 5GHz}$ bins, 
independently on the classification originally attached to each source.  
The result is shown in Fig.~1.
Some trends (see also Sambruna et al. 1996b) are evident: 

\smallskip
\item{$\bullet$} the {\it X--ray spectrum} becomes harder while 
the $\gamma${\it --ray spectrum} softens with increasing luminosity 
(see also Comastri et al. 1997).

\item{$\bullet$} {\it two peaks} are present in all the SEDs. 
The first peak moves from $\sim 10^{16}~-~10^{17}$ Hz for less luminous
sources, to $\sim 10^{13}~-~10^{14}$ Hz for the most luminous ones.
The location of the second peak can be inferred from the shape of the 
$\gamma$--ray spectrum ($\gamma$--ray spectrum steep/medium/flat 
$\rightarrow$ peak below/within/above the EGRET energy range).

\item{$\bullet$} for a given SED {\it the frequencies of the two peaks are 
correlated}.
 
\item{$\bullet$} the {\it strength of the $\gamma$--ray peak} with respect 
to the lower frequency one seems to increase with increasing L$_{\rm 5GHz}$ 

\smallskip

\noindent
These trends are also present in
each of the samples, and strengthened when the total blazar sample is
considered, suggesting that we have to do with a {\it continuous spectral 
sequence} within the blazar family, rather than with separate 
spectral classes.

\topinsert
\vglue -0.6 cm
\centerline{\epsfxsize=12.5 cm \epsfbox{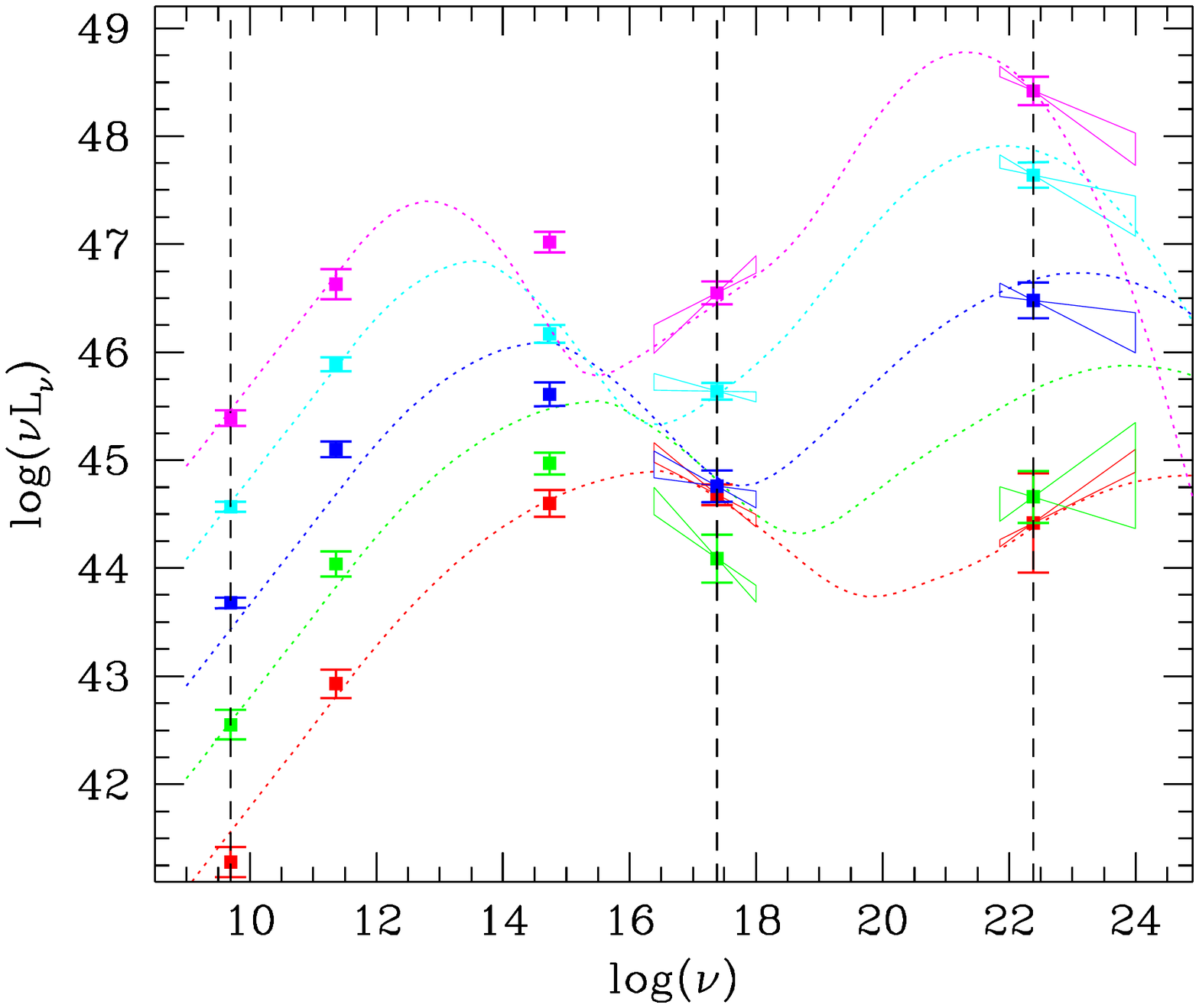}}
\vskip -3.5 truecm
\setbox0=\vtop{
{\figure{3}{``total'' blazars sample average SEDs with analytic SEDs
superimposed.}}
}
\line{\box0}
\endinsert

In Fig.~1 we superimposed to real data a set of (dotted) lines, whose
main goal is to guide the eye.
However we note that the radio--X--ray part is the analytic parameterization
proposed by Fossati et al. (1997a), based on the hypothesis that the
peak frequency of the first spectral component is inversely related 
to its luminosity ($\nu_{peak} \propto L^{-1.5}$ in this case). 

The second spectral component has been derived assuming that 
(a) the ratio of the frequencies of the high and low energy peaks is constant 
($\nu_H/\nu_L~\simeq 5 \times 10^8$), 
(b) the high energy peak and radio intensities have a fixed ratio.
Given the extreme simplicity of these assumptions, it is remarkable that 
the phenomenological model fits reasonably well the average SEDs.

\titleb{4}{Conclusions}

The main conclusion is that {\it despite/because of} clear systematic
differences between the continua of different blazar sub--classes,
a unitary approach is possible.

The physical reason(s) underlying the systematic trends in the
SEDs of blazars illustrated here are at present unknown.
Ideas under discussion involve the effect of radiative losses 
on the particle distribution, which could lead to a break at a 
critical energy $\gamma_{e,break}$. In this
case in fact, $\gamma_{e,break}$ would inversely depend on   
the compactness in seed photons for inverse Compton losses,
i.e. $\gamma_{e,break} \propto U^{-1}_{rad}$) 
(Comastri et al. 1997, Fossati et al. 1997a, Ghisellini et al. 1997). 

Systematic determinations of the physical parameters for a number of objects
with different values of $\nu_{peak, 1, 2}$
can therefore give important clues on the relationship between BL Lac objects
and FSRQs and on the physical processes governing the radiative properties
of relativistic jets (Ghisellini et al. 1997).

\begrefchapter{References}

\ref Comastri, A., Fossati, G., Ghisellini, G., \& Molendi, S., 1997, ApJ, 480, 534 

\ref Fossati, G., Celotti, A., Ghisellini, G., \& Maraschi L., 1997a, MNRAS., in press.

\ref Fossati, G., Celotti, A., Comastri, A., Ghisellini, G., \& Maraschi L., 1997b, in preparation

\ref Ghisellini, G., Celotti, A., Comastri, A., Fossati, G., Maraschi, L., 1997, in preparation

\ref Impey, C. 1996, AJ, 112, 2667

\ref Lamer, G., Brunner, H., Staubert, R. 1996, A\&A, 311, 384

\ref Mattox, J.R., et al. 1997, ApJ, 481, 95

\ref Padovani, P., and Urry, C.M. 1992, ApJ, 387, 449 

\ref Perlman, E., et al. 1996a ApJS, 104, 251

\ref Perlman, E., et al. 1996b ApJ, 456, 451

\ref Sambruna, R.M., et al. 1996a, ApJ, 463, 424

\ref Sambruna, R.M., Maraschi, L., \&  Urry, C. M. 1996b, ApJ, 463, 444

\ref Sambruna, R.M. 1997, ApJ submitted

\ref Stickel, M. et al. 1991, ApJ, 374, 431

\ref Wall, J.V., and Peacock, J.A. 1985, MNRAS, 216, 173

\endref

\byebye